\documentclass[conference]{IEEEtran}

\usepackage{graphicx} 
\usepackage{subfigure}

\usepackage[T1]{fontenc}
\usepackage{amsmath}
\usepackage{amssymb}
\usepackage{widetext}

\newcommand{\weight}{w_\beta}

\newcommand{\risk}{R}
\newcommand{\cb}{c^\bot}
\newcommand{\Csetg}{\mathcal{Z}}
\newcommand{\dC}{\Csetg_{12}}

\newcommand{\rate}{\rho}
\newcommand{\trans}{\tau}

\newcommand{\wSums}{weight sums }
\newcommand{\jwSum}{joint weight sum }

\begin{document}
\title{Selecting the rank of truncated SVD\\ by Maximum Approximation Capacity}
\author{\IEEEauthorblockN{ Mario Frank 
 and
Joachim M. Buhmann
}
\IEEEauthorblockA{ 
mario.frank@inf.ethz.ch,
 jbuhmann@inf.ethz.ch}
\IEEEauthorblockA{ Department of Computer Science, ETH Zurich;  Universitätsstrasse 6, CH-8006 Zürich, Switzerland }
}

\maketitle
\begin{abstract}
Truncated Singular Value Decomposition (SVD) calculates the closest
rank-$k$ approximation of a given input matrix. Selecting the
appropriate rank $k$ defines a critical model order choice in most
applications of SVD. To obtain a principled cut-off criterion for the
spectrum, we convert the underlying optimization problem into a
noisy channel coding problem. The optimal approximation capacity of
this channel controls the appropriate strength of regularization to
suppress noise. In simulation experiments, this information theoretic
method to determine the optimal rank competes with state-of-the art
model selection techniques.
\end{abstract}
\IEEEpeerreviewmaketitle

\section{Introduction}
Singular Value Decomposition (SVD) is a widely used 
technique for exploratory data analysis. It decomposes a given input
matrix into a product of three matrices such that
$\mathbf{X}=\mathbf{U}\mathbf{S}\mathbf{V}^T$. Thereby, $\mathbf{U}$
and $\mathbf{V}$ are unitary matrices and they essentially induce a
rotation of the input data. $\mathbf{S}$ is a diagonal matrix
(inducing a scaling) with the singular values as entries. Quite often,
one is rather interested in an approximation of $\mathbf{X}$ instead
of the exact decomposition such as, for instance, in Principal
Components Analysis (PCA). For SVD, this approximation technically
requires to set all but the first $k$ diagonal entries in $\mathbf{S}$
to $0$. The resulting reconstruction $\mathbf{U}\mathbf{S}_k\mathbf{V}^T$ has rank
$k$. Neglecting all but the first $k$ components is justified since
the noise in the data perturbs the small eigenvalues, whereas the first $k$
components supposedly capture the underlying structure or the signal
of the data. Selecting the cutoff value $k$ defines the central
model-order selection problem of truncated SVD. 

In this paper we propose a novel method for selecting $k$ which is
based on the framework of approximation set coding
(ASC) \cite{ISIT2010}. ASC defines the maximum approximation capacity
(maxAC) principle for model-order selection.
 For a
given model order and a given noisy dataset, ASC theory enables us to
compute the capacity of a hypothetical channel. MaxAC selects the
model that achieves highest capacity, i.e., the model of highest
complexity that still can be robustly optimized in the presence of
noise. Originally, maxAC was derived for discrete optimization
problems. So far, it was applied to decoding for the binary channel
\cite{ISIT2010} and for selecting the number of clusters in clustering problems \cite{ASC4clustering}. In
this contribution, we adopt it, for the first time, to a continuous
optimization problem, namely for selecting the cut-off rank of truncated SVD. Thereby, we investigate the challenges of the continuous solution
space. 
Moreover, we provide an
experimental comparison with other model-selection methods indicating
that our maxAC adaption to SVD can compete with state-of-the art
methods. 
 In the remainder of the paper we will first introduce the
main concepts of ASC in Section~\ref{sec_theory}. In Section~\ref{sec_asc4svd} we then derive the approximation capacity for SVD, point to problems and
solutions and report our numerical findings.

\section{Approximation Set Coding \label{sec_theory}}
In this section we briefly recapitulate the principle of maximum
approximation capacity (maxAC). Due to the page limit, we must refer
to the original paper \cite{ISIT2010} for a more detailed description.

Consider a generic optimization problem with the input $\mathbf{X} =
\{\mathbf{x}_1,\dots,\mathbf{x}_N\} \in \mathcal{X}$ consisting out of
$N$ measurements. The vector $\mathbf{x}_i\in\mathbb{R}^D$ identifies
the $i^{\textrm{th}}$ measurement. Let the output of the optimization
problem be the \textbf{solution} or \textbf{hypothesis} $c\in
\mathcal{C}$ where $ \mathcal{C}$ is the set of all hypotheses
satisfying the constraints of the problem, the \textbf{hypothesis
  space}.  An optimization problem involves a cost function
$\risk\,:\,\mathcal{C} \times \mathcal{X} \rightarrow \mathbb{R}_{\ge
  0}$ that maps a hypothesis $c$ to a real value
$\risk(c,\mathbf{X})$.  Finally, let $\cb = \arg\min_{c}
\risk(c,\mathbf{X})$ be the hypothesis that minimizes costs.  In order
to rank all solutions of the optimizer, we introduce
\emph{approximation weights}
\begin{eqnarray}
  w :  \mathcal{C}\times \mathcal{X}\times \mathbb{R}_+ &\rightarrow&
  [0,1]\,, \\
  (c,\mathbf{X},\beta) &\mapsto& \weight(c,\mathbf{X}) = \exp(- \beta
\risk(c,\mathbf{X})) \label{eq_weights} \,. \nonumber
\end{eqnarray}
These weights, parameterized by the inverse computational temperature
$\beta$, define the two \emph{\wSums} $\Csetg_{1}, \Csetg_{2}$, and the
\emph{\jwSum} $\dC$
\begin{eqnarray}
\label{eq:Bfac_approx1}
\Csetg_{\nu} &:=&  \sum_{c\in\mathcal{C}}
        \exp\left(- \beta    \risk(c,\mathbf{X}^{(\nu)}) \right),\: \nu=1,2 \\
\label{eq:Bfac_approx2}
 \dC &:=&
 \sum_{c\in\mathcal{C}} 
    \exp \left(- \beta (\risk(c,\mathbf{X}^{(1)})+ \risk(c,\mathbf{X}^{(2)})) \right)    ,
\end{eqnarray}
where $\mathbf{X}^{(1)}$ and $\mathbf{X}^{(2)}$ are two random subsets
of the input $\mathbf{X}$ and $\exp(- \beta
(\risk(c,\mathbf{X}^{(1)})+ \risk(c,\mathbf{X}^{(2)})))$ measures how
well a single solution $c$ minimizes costs on both datasets. The sums
(\ref{eq:Bfac_approx1},\ref{eq:Bfac_approx2}) play a central role in
ASC. If $\beta=0$, all weights $ \weight(c,\mathbf{X}) =1$ are
independent of the costs. In this case, $\Csetg_{\nu} = |\mathcal{C}|$
indicates the size of the hypothesis space, and
$\dC=\Csetg_1=\Csetg_2$.  For intermediate $\beta$, $\Csetg(\cdot)$
takes a value between $0$ and $|\mathcal{C}|$, giving rise to the
interpretation of $\Csetg(\cdot)$ as the effective number of
hypotheses that approximately fit the dataset $\mathbf{X}^{(\nu)}$,
whereas $\beta$ defines the precision of this approximation.

ASC constructs a hypothetical coding scenario where sender and
receiver have to communicate problem solutions via transformations
$\trans \in \mathcal{T}$ of noisy datasets $\mathbf{X}^{(1)}$ and
$\mathbf{X}^{(2)}$. As derived in \cite{ISIT2010}, an asymptotically
vanishing error rate for such a channel is achievable for rates
\begin{eqnarray}\label{mInfo}
  \rate &\leq&
\mathcal{I}_\beta(\trans_s,\hat{\trans}) 
= {\frac{1}{N}}
    \log\left(
    \frac{
         \vert\{\trans_s\}\vert \cdot \dC
    }{
        \Csetg_1 \cdot \Csetg_2
    }
    \right) \ .
\end{eqnarray}
Eq.(\ref{mInfo}) denotes the mutual information between the encoded
transformation $\trans_s$ of the dataset and the decoded
transformation $\hat{\trans}$. A higher approximation precision $\beta$
has two effects. First, the hypothesis space is covered by more
patches of approximate solutions leading to more available codewords
and a higher codeword entropy $\vert\{\trans_s\}\vert$. Second, these
patches have a higher overlap such that, in the decoding step, codewords are possibly 
confused. The maximum of
$\mathcal{I}_\beta(\trans_s,\hat{\trans}) $ with respect to $\beta$ is
called the approximation capacity.

\section{Approximation capacity of SVD}
\label{sec_asc4svd}
To select the cut-off rank for SVD, we have to find the rank $k^*$ with
maximal approximation capacity. This search involves maximizing
Eq.(\ref{mInfo}) for all ranks that we investigate. By going through
the list of all quantities introduced in the last section, we relate
them to truncated SVD. The input of the problem is a matrix
$\mathbf{X}\in \mathbb{R}^{N \times D}$. The cost function is the
Frobenius norm:
\begin{equation}
\label{eq_frob}
R(\mathbf{X},\mathbf{U},\mathbf{S},\mathbf{V} ) = \sum_{i,j} \left(
x_{ij} - \sum_{t=1}^k u_{it}s_{tt}v_{tj} \right)^2
\end{equation}
An optimizer for this problem outputs a decomposition $
\cb(\mathbf{X}^{(\nu)})=\mathbf{U}^{(\nu)}\mathbf{S}^{(\nu)}\mathbf{V}^{(\nu)}
$, whereas all but the first $k\leq \min(N,D)$ diagonal entries of
$\mathbf{S}^{(\nu)}$ are $0$ due to truncation (we will denote the
third matrix by $\mathbf{V}$ instead of $\mathbf{V}^T$ for
convenience).  The solution $\cb(\mathbf{X}^{(\nu)})$ gives the closest rank-$k$
approximation of $\mathbf{X}^{(\nu)}$ with respect to the Frobenius
norm. In the case of SVD, a hypothesis
$c=(\mathbf{U},\mathbf{S},\mathbf{V})$ is a particular decomposition
of the input matrix (in clustering, for instance, $c$ is a relation
that assigns objects to clusters \cite{ASC4clustering}). The $k\times
D$ matrix with the entries $w_{tj}:=s_{tt}v_{tj}$ is a new basis and
$\mathbf{U}$ provides the linear weights needed to represent the data
$\mathbf{X}$ in this basis. When the empirical mean of $\mathbf{X}$ is
the origin, this representation corresponds to PCA.

We define the hypothesis space for truncated SVD with cut-off rank $k$
as follows.  For a fixed basis $\mathbf{W}$, the hypothesis space
 is spanned by all $N\times k$ matrices $\mathbf{U}$. For
a given dataset $\mathbf{X}^{(\nu)}$, $\mathbf{U}^{(\nu)}$ is the
cost-minimizing solution. We parameterize the different
transformations $\trans\in\mathcal{T}$ for encoding messages in
datasets by $\trans\circ\mathbf{X}^{(\nu)}=\mathbf{X}^{(\nu)} +
\mathbf{U}_{\trans}\mathbf{W}^{(\nu)}$. The identity transformation is
$\mathbf{U}_{\text{id}}=0$.

The mutual information Eq.~(\ref{mInfo}) was 
originally derived for finite hypothesis spaces, such as for
clustering solutions or binary message strings.  The challenges of
computing the weight sums (\ref{eq:Bfac_approx1}) and
(\ref{eq:Bfac_approx2}) are twofold.  First, in a small volume of a
continuous hypothesis space, there are infinitely many
transformations.  Second, transformations can have infinite distance
to each other such that the receiver can always distinguish an
infinite subset of $\mathcal{T}$ even when the datasets are
noisy. This makes the union bound in the derivation of the
error-probability in \cite{ISIT2010}, Eq.~(7) inadequate.  For these
reasons, calculating the capacity under the assumption that
$\mathbf{U}$ can be any real $N\times k$ matrix fails. In the
following, we will first demonstrate the effect of this assumption by
providing the na\"{\i}ve analytical calculation of the mutual
information in Eq.~(\ref{mInfo}). Then we introduce constraints on the transformations
$\mathcal{T}$ such that Eq.~(\ref{mInfo}) can be computed.

\subsection{Unconstraint Hypothesis Space}
\label{subsec_analyt}
We abbreviate
$R_{\nu}=R(\mathbf{X}^{(\nu)},\mathbf{U},\mathbf{S}^{(\nu)},\mathbf{V}^{(\nu)}
)$ and $R_{\Delta}= 1/2 \left(
R(\mathbf{X}^{(1)},\mathbf{U},\mathbf{S}^{(1)},\mathbf{V}^{(1)} ) +
R(\mathbf{X}^{(2)},\mathbf{U},\mathbf{S}^{(1)},\mathbf{V}^{(1)}
)\right)$ and use the short-hand notation of $\mathbf{m_{i,.}}$ as the $i^{\text{th}}$ row of a matrix $\mathbf{M}$.  By integrating over the space of all linear combinations
$\mathbf{U}$, we obtain the weight sums:
\begin{align}
&\Csetg_{\nu} \!= \!\!\int_{-\infty}^{\infty}\!\!\!\exp\left[ - \beta
    R_{\nu} \right] {d\mathbf{U}} \ ,\ \ \nu\in\left\{1,2\right\} 
\\
&=\prod_{ij}e^{-\beta {x_{ij}^{(\nu)}}^2 }  
\!\!\int_{-\infty}^{\infty}\!\!\! e^{ -\beta \left(\left(\sum_t^k
  u_{it}w_{tj}^{(\nu)}\right)^2 -2x_{ij}^{(\nu)}\sum_t^k
  u_{it}w_{tj}^{(\nu)} \right) } {d^k\mathbf{u_{i,.}}}  
\nonumber
\\
&\dC \!= \!\!\int_{-\infty}^{\infty}\!\!\!\exp\left[ -\beta R_{\Delta} \right] {d\mathbf{U}} =  \prod_{ij}e^{-\beta/2 \left({x_{ij}^{(1)}}^2 +{x_{ij}^{(2)}}^2 \right) }
\\
&\cdot\!\!\int_{-\infty}^{\infty}\!\!\! e^{ -\beta/2 \left( 2\left(\sum_t^k u_{it}w_{tj}^{(1)}\right)^2 
-2\left(x_{ij}^{(1)}+ x_{ij}^{(2)}\right)\sum_t^k u_{it}w_{tj}^{(1)} \right) } {d^k\mathbf{u_{i,.}}} 
\nonumber
\end{align}
The solution of Gaussian integrals yields
\begin{eqnarray*}
\!\!\int_{-\infty}^{\infty}\!\!\! e^{-1/2 \sum_{t}^k \sum_{t'}^k A_{t,t'} u_t u_{t'} +\sum_{t}^k b_t u_t }
\!\!&\!\!=\!\!&\!\!\!\!
\sqrt{\frac{\left(2\pi\right)^k}{\det(\mathbf{A})}}
e^{1/2 \mathbf{b}^T\mathbf{A}^{-1}\mathbf{b}}  \ ,
\end{eqnarray*}
whereas in our case 
\begin{align}
&\mathbf{A^{(\nu)}_{(j)}} \!\!=\!\! 2\beta \;\mathbf{w_{.,j}^{(\nu)} }^T\mathbf{w_{.,j}^{(\nu)} } \ , \quad \ 
\mathbf{b^{(\nu)}_{(j)}} \!\!=\!\! 2\beta \;x_{ij}^{(\nu)}\mathbf{w_{.,j}^{(\nu)} }^T \ ,
\\
&\mathbf{A^{(\Delta)}_{(j)}} \!\!=\!\! \mathbf{A^{(1)}_{(j)}} \ , \quad \quad \quad \quad \quad
\mathbf{b^{(\Delta)}_{(j)}} \!\!=\!\! \beta \left(x_{ij}^{(1)}+x_{ij}^{(2)}\right)\mathbf{w_{.,j}^{(1)} }^T \ . \nonumber
\end{align}
The weight sums are
\begin{align}
& \Csetg_{\nu} \!\!  =\!\!  \prod_{i,j} e^{-\beta x_{ij}^{(\nu)2}} \!\! 
\sqrt{ \frac{(2\pi)^k}{\det\left(2\beta \;\mathbf{w_{.,j}^{(\nu)} }^T \mathbf{w_{.,j}^{(\nu)} } \right) }   }  
 e^{1/2 \mathbf{b}^T\mathbf{A}^{-1}\mathbf{b}}
\nonumber
\\
&\stackrel{(i)}{=}\!\!
\left(\frac{\pi}{\beta}\right)^{\!\!\frac{NDk}{2}} 
 \!\!\!\!\!\!\prod_j \!\!\left(\!D^{(\nu)}_j\! \right)^{\,\frac{\!-N}{2}} \!\!
e^{\!-\beta  \sum_{\!ij} \left(x_{ij}^{(\nu)2} \left(1- \mathbf{w_{.,j}^{(\nu)} }^T (\mathbf{w_{.,j}^{(\nu)} } \mathbf{w_{.,j}^{(\nu)} }^T)^{-1}\mathbf{w_{.,j}^{(\nu)} } \right) \right)  }
\nonumber
\\
 &\stackrel{(ii)}{=}\!\!
\left(\frac{\pi}{\beta}\right)^{\frac{NDk}{2}} 
\prod_j\left( D^{(\nu)}_j \right)^{\,\frac{-N}{2}}
  \\
& \dC = 
  \left| C^{(1)}_{\gamma}\right|
\prod_{ij} e^{-\beta /4 \left(x_{ij}^{(1)} -x_{ij}^{(2)}\right)^2}\ .
\end{align}
We abbreviated $D^{(\nu)}_j :=\det(\mathbf{w_{.,j}^{(\nu)T}  } \mathbf{w_{.,j}^{(\nu)} } )$.
In step (i), we used that for a $n\times n$ matrix $\mathbf{M}$ and a scalar $p$, it holds that $\det\left( p \mathbf{M} \right) = p^n\det\left(\mathbf{M} \right) $.
In step (ii), we used that $\mathbf{F^{(\nu)}} :=\mathbf{w_{.,j}^{(\nu)} } (\mathbf{w_{.,j}^{(\nu)} }^T \mathbf{w_{.,j}^{(\nu)} } )^{-1}\mathbf{w_{.,j}^{(\nu)} }^T=1$.
Substituting to Eq.~(\ref{mInfo}) provides the mutual information:
\begin{align}
\label{mInfo_analyt}
&I(\beta) 
\!= 
\frac{1}{N} \left( \log \left(\dC\right)
- \log\left( \Csetg_{1} \right) - \log\left( \Csetg_{2} \right) \right) 
\\
&=
\frac{Dk}{2} \log\left(\frac{\beta}{\pi}\right)
+\!\! \frac{1}{2} \sum_j D^{(2)}_j \!\!
-\frac{\beta}{4N} \sum_{ij} \left(x_{ij}^{(1)} -x_{ij}^{(2)}\right)^2  \ . \nonumber
\end{align}
The first order condition provides the optimal temperature
\begin{align}
1/\beta^* = \frac{1}{2NDk} \sum_{ij} \left(x_{ij}^{(1)} -x_{ij}^{(2)}\right)^2 \ .
\end{align}
Note that the temperature monotonically increases with the distance of
the two datasets, as one would expect: with more noise, the precision
for approximating the optimal solution must be lower. However,
unexpectedly, the temperature decreases with $k$ suggesting that a
higher rank stabilizes the solutions. This misconception is a
consequence of the unconstrained hypothesis space, as discussed earlier,
and indicates that constraints for $\mathbf{U}$ are necessary.  Also,
we neglected the temperature-independent term $\vert\{\trans_s\}\vert$
in Eq.~(\ref{mInfo}) which would be infinitely high.

\begin{figure*}[htb]
\centering
\includegraphics[width=0.24\textwidth]{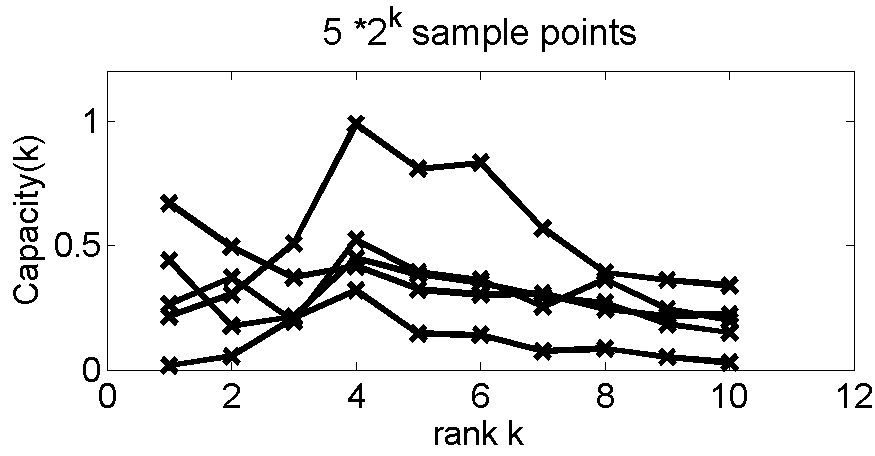}    
\includegraphics[width=0.24\textwidth]{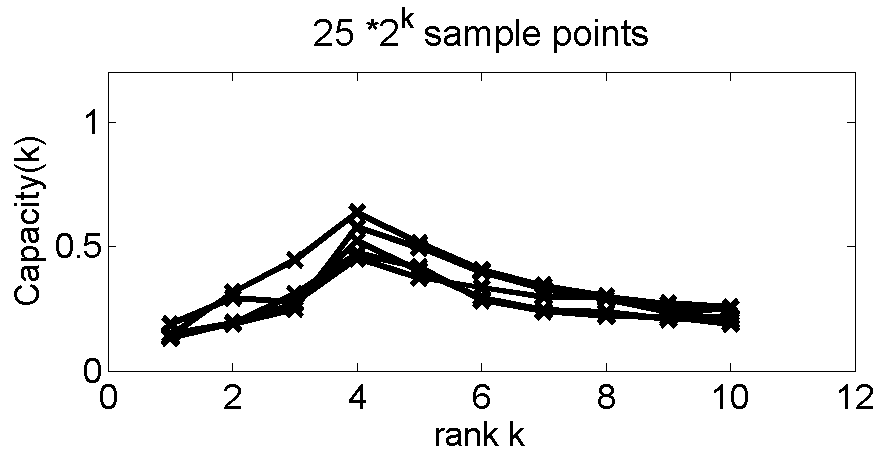}
\includegraphics[width=0.24\textwidth]{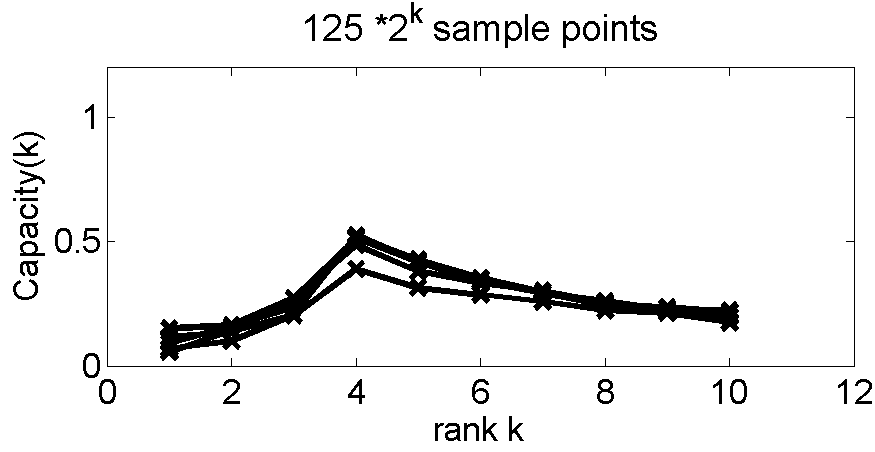}
\includegraphics[width=0.24\textwidth]{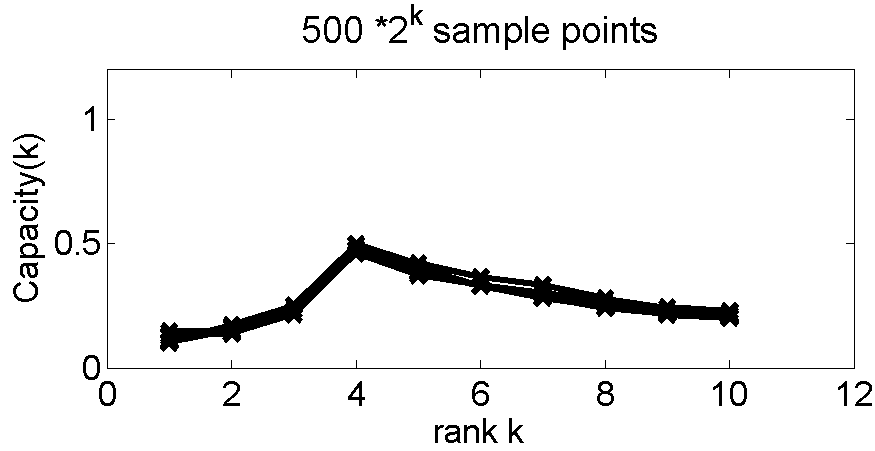} 
\caption{ Approximation capacity against rank for various numbers of
  sample points. The optimal rank is $k=4$.  Even though the
  individual trends still vary a lot for very few sample points, the
  optimal model-order is already found. With increasing number of
  transformations the calculations are stabilized.
\label{fig:infOfRes}  }
\end{figure*}

\subsection{Finite and Bounded Hypothesis space} 
In the discussion above, we identified two problems: an infinitely
large capacity due to i) an infinitely large transformation space
$\mathbb{R}^{N\times k}$ (or a negative one if we disregard the
infinitely many possible codewords) and ii) due to the existence of
infinitely many transformations in an arbitrary small volume of this
space. For a practical implementation of the maxAC criterion for SVD
we have to i) bound the hypothesis space and ii) constraint the
density of transformations to a finite number. This renders the
integrals for computing the weight sums to finite sums which must be
explicitly computed. In our experiments, we use two ways of summing
over the hypothesis space. First, the transformations populate the
hypothesis space on an equispaced grid in a hypercube of finite
size. Second, we randomly sample transformations from an isotropic
Gaussian. In both cases the set of transformations is centered around
the cost minimizing solution $\mathbf{U}^{(1)}$ (the identity
transformation $\mathbf{U}_{\text{id}}$). For both ways, one must
choose the boundaries (the size of the grid or the variance of the
Gaussian) as well as the number of transformations.

\begin{figure}[htb]
\centering
\includegraphics[width=0.33\textwidth]{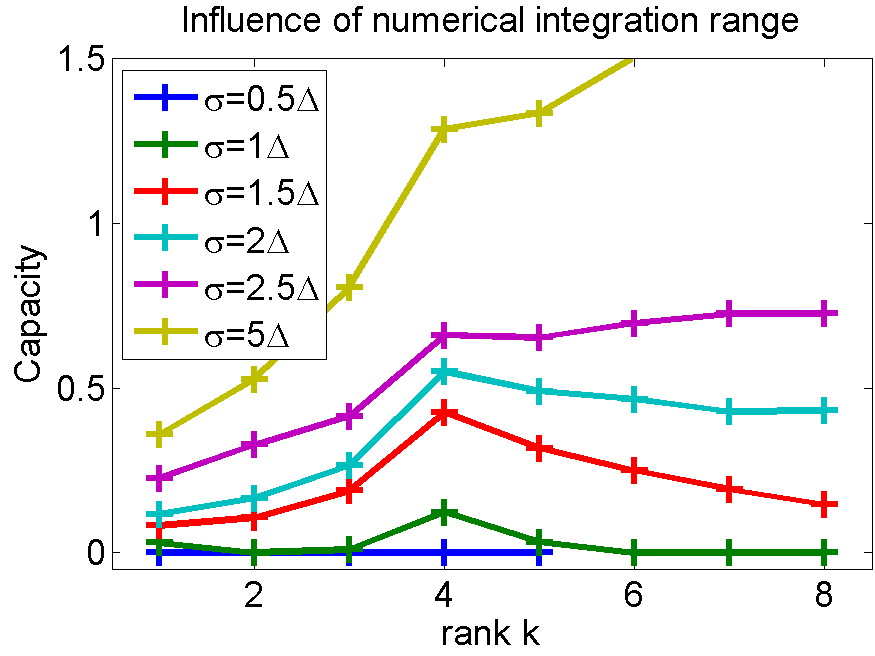} 
\caption{Numerically computed approximation capacity for various sizes
  of the subspace containing the transformations.\label{fig:infOfRange} }
\end{figure}

We experimentally investigate the influence of these parameters.
First, we study the influence of the integration range on the
capacity. We create data $\mathbf{X}^{(\nu)}$ from a mixture of 4
Gaussians with isotropic noise, leading to an optimal rank 4. We
compute the approximation capacity by sampling transformations from a
Gaussian sphere around the cost-minimizing SVD solution
$\mathbf{U}^{(\nu)}$ with standard deviation $\sigma$. Our
experimental findings for various magnitudes of $\sigma$ are
illustrated in Fig.~\ref{fig:infOfRange}. We write $\sigma$ in units
of $\Delta = 1/N \sum_{i=1}^N
\left\|\mathbf{u}_i^{(1,2)}-\mathbf{u}_i^{(1)}\right\|$, whereas
$\mathbf{U}^{(1,2)}$ is the matrix that satisfies
$\mathbf{X}^{(2)}=\mathbf{U}^{(1,2)}\mathbf{W}^{(1)}$. In the regime
where $1/N\left\|\mathbf{U}_{\trans} \right\|\approx \Delta $, a
transformed dataset $\trans_j \circ \mathbf{X}^{(2)}$ could possibly
be confused with $\mathbf{X}^{(1)}$.  When the transformations are
smaller than $\Delta $, none of the transformations could possibly be
used in a codebook as they are all indistinguishable from the identity
transformation. As a result, the obtained capacity is too low.  On the
contrary, for a too high integration range, the capacity converges to
the na\"{\i}ve analytical solution because infinitely many
transformations could serve as distinguishable codewords.

The second experiment studies the influence of the number of
transformations on the mutual information. This time, we use a grid of
fixed size and vary the density of grid points. In order to
sufficiently cover the hypothesis space, we increase the number of
transformations by a factor of 2 when increasing $k$. While this increment is still too low to preserve the transformation density, it already imposes a computational challenge. In our experiments with
larger datasets, we sample the hypothesis space more sparsely.  The influence of the number of transformations is illustrated in Fig.~\ref{fig:infOfRes}. The results demonstrate that this  number only affects the stability of the computation and not the maximum of the capacity.

\subsection{Continuous and bounded hypothesis space} 
The numerical experiment on the influence of the number of transformations suggests that for a defined transformation density, the analytical solution should provide the desired result if only the integration range is properly defined. We calculate the mutual information as in Section~\ref{subsec_analyt} but, this time, we weight the integrand with an isotropic Gaussian around the identity $u^{(1)}_{it},\forall \:i,t$ to suppress the contribution of heavy transformations. 
\begin{align}
 \Csetg_{\nu}  \!&=\!\!  \int_{-\infty}^{\infty}\!\!\!e^{ -\beta R_{\nu} }  e^{-\frac{1}{2\sigma}\sum_{i,t} (u_{it}-u^*_{it})^2}  {d\mathbf{U}} ,\ \nu\in\left\{1,2\right\}
\\
\dC \!&=\!\!  \int_{-\infty}^{\infty}\!\!\!e^{ -\beta R_{\Delta} }  e^{-\frac{1}{2\sigma}\sum_{i,t} (u_{it}-u^*_{it})^2}  {d\mathbf{U}}
\end{align}
The derivation of the mutual information is provided in Appendix B.
The simulations depicted in Fig.~\ref{fig:infOfRange_analyt} illustrate that for an analytically computed mutual information (see Eq.~(\ref{mInfo_analyt_bounded})), the 
width $\sigma$ influences the capacity much more than in the numerical computation (compare with Fig.~\ref{fig:infOfRange_analyt}). However, if a maximum exists (square markers in Fig.~\ref{fig:infOfRange_analyt}), it is at the correct rank.

\begin{figure}[htb]
\centering
\includegraphics[width=0.33\textwidth]{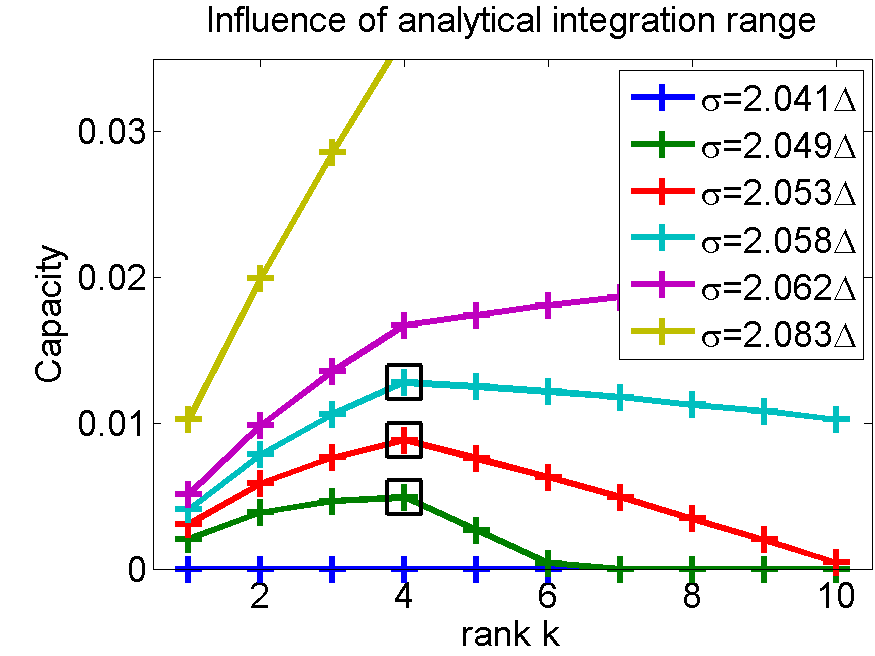}
\caption{Analytically computed approximation capacity integrated over an isotropic Gaussian sphere with varying standard deviation.\label{fig:infOfRange_analyt} } 
\end{figure}

\subsection{Comparison with other model-selection techniques}  
We study how well maxAC and other model-order selection methods select
the appropriate rank for approximating a noisy dataset via
rank-limited SVD. 
A comparison with the following methods is performed: 
'Laplace'
and 'BIC' approximate the marginal likelihood (the evidence) for
probabilistic PCA \cite{Minka00automaticchoice}. The first method
applies Laplace approximation to the involved integral. The well-known
BIC score \cite{bic} further simplifies that the likelihood exhibits
the same sensitivity to all model parameters. A third method, the
minimum transfer cost principle ('MTC') \cite{mtc_ECML}, mimics
cross-validation. It learns a
rank-limited SVD on one random subset of the data and then computes
the costs when applying it to another one. Thereby, the model
parameters are transfered by a mapping function $\psi$ which maps each
object in the first dataset to its nearest neighbor in the second
dataset.

We create objects from a number of centroids with a defined separation
from each other and add Gaussian noise. The difficulty of the problem
is controlled by altering the variance relative to the separation of
the centroids.  When going to a higher number of centroids, the
dimensionality must also be increased to preserve their
separation. To enable a comparison with the PCA methods, the
data mean is shifted to the origin.
 
\begin{figure*}[htb]
\centering
\subfigure[Low relative noise]{\includegraphics[width=0.31\textwidth]{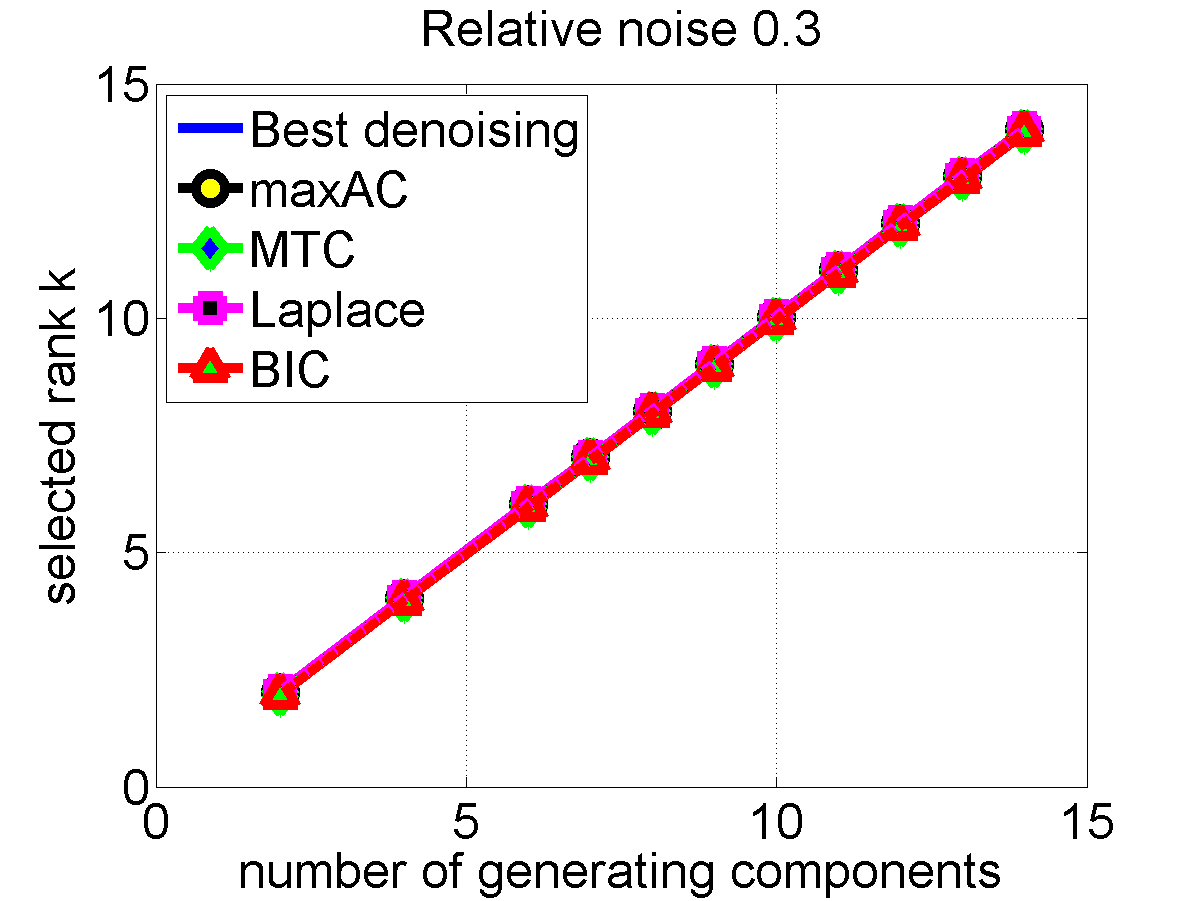} \label{subfi_lonoise}}
\subfigure[High relative noise]{\includegraphics[width=0.31\textwidth]{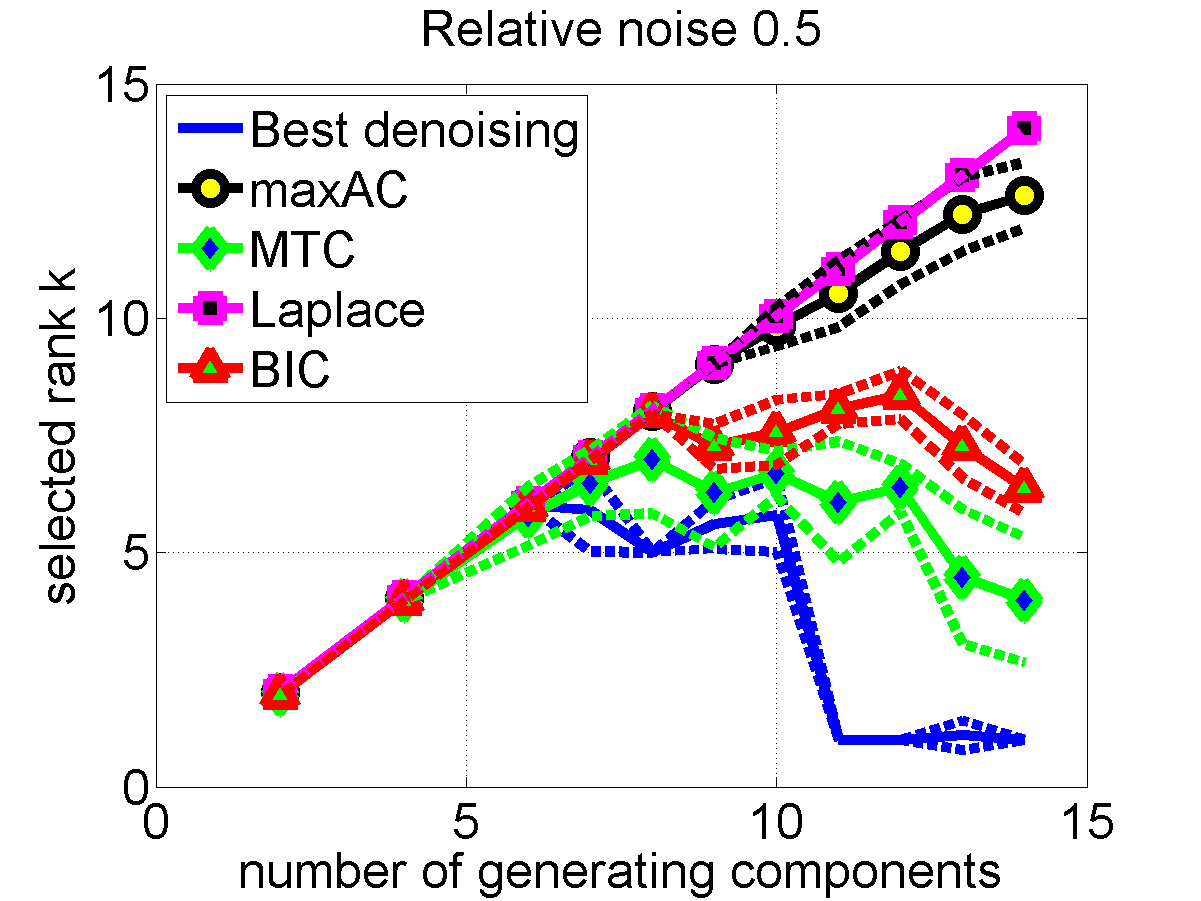} \label{subfi_hinoise}}
\subfigure[Various noise levels]{\includegraphics[width=0.31\textwidth]{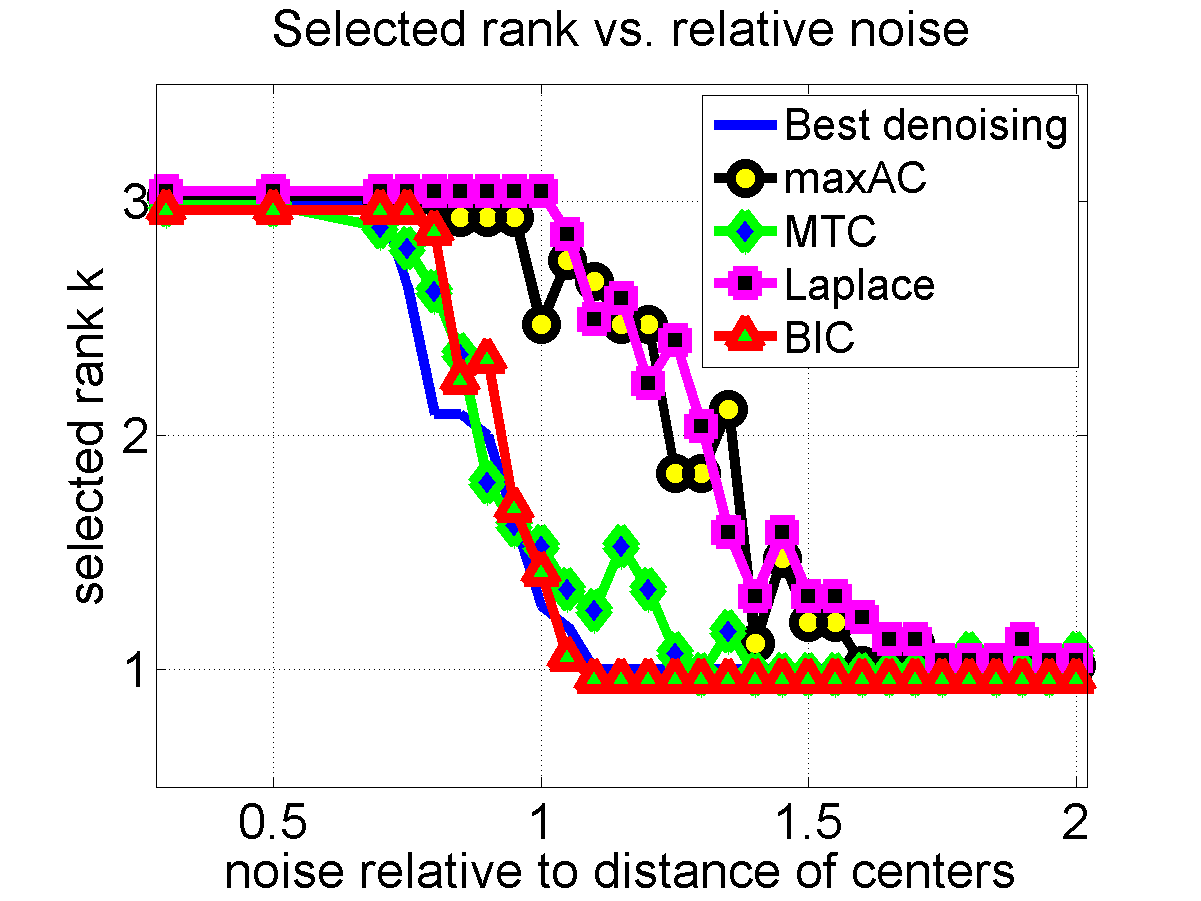}
\label{subfi_varoise}} 
\caption{Rank-k approximation for a mixture of Gaussians. In (a) and (b) the number of generating components varies, in (c) the noise level varies for 3 components. For readability, we omitted the variances in (c). They are comparable with (b).
All methods select a rank between the true number of components and the rank that minimizes the distance to the noise-free matrix (`Best denoising'). \label{fig:modSelGaussian}  }
\end{figure*}

For a given true number of generating components and a given noise
level relative to the centroid separation, there exists one SVD rank
that yields the reconstruction with the minimal deviation from the
noise-free matrix. For very noisy data or a high number of components
and dimensions, this optimally denoising rank is smaller than the true
number of generating components. Inspecting
Fig.~\ref{fig:modSelGaussian}, one can see that all methods select a
rank between the best denoising rank and the true rank. For low noise
(Fig.~\ref{subfi_lonoise}) learning the rank is easy. For a high
noise level, the learning problem becomes hard when the number of generating components and the dimensionality increase. 
There exists a
transitional regime where all methods start selecting a lower rank
than the true one (Fig.~\ref{subfi_hinoise}). 
For very high noise levels, all methods select $k=1$ (Fig.~\ref{subfi_varoise}).

\section{Conclusion}
We proposed a novel technique for selecting the cut-off rank of truncated SVD. Our criterion selects the rank that
maximizes the approximation capacity (maxAC) and, thereby, captures
the maximum amount of information in the data that can by reliably
inferred from random subsets of the input dataset. We demonstrated, for the first time, how to apply
maxAC to an optimization problem with a continuous solution space like
SVD. The Euclidean geometry of the parameter space renders the computation of the approximation capacity more difficult than in the clustering case with random
permutations \cite{ASC4clustering}. We discussed the challenges with such
problem domains and proposed solutions. Finally, we demonstrated in
comparative experiments that the model-order selection of our technique conforms with established methods.  Future work will address the
application of the maxAC criterion to other continuous optimization
problems such as, for instance, sparse linear regression.

\section*{Acknowledgments}
This work was partially supported by the Zurich Information Security Center and by
the FP7 EU project SIMBAD.

\bibliographystyle{IEEEtran}
\bibliography{biblio}

\newpage

 \section*{Appendix A: attaining capacity} 
 There are several ways of numerically maximizing the mutual information in Eq.~{\ref{mInfo}} with respect to the temperature. The simplest way is to compute $I_\beta$ for several values of $\beta$ and pick the maximum. This creates a quantization error of the optimum that depends on the step-size of the temperature scale. Moderate step-sizes already lead to good results as $I_\beta$ is usually flat around its maximum. 
 
 Using a gradient-descent method like Newton iterations provide more precise results .
In the following, we report the first and the second derivation of $I_\beta$ which are needed for the Newton updates.
The derivation of the mutual information $I_\beta$ (Eq.~{\ref{mInfo}}) with respect to the inverse temperature $\beta$ is
\begin{align}
&\frac{\partial I(\beta)}{\partial \beta} 
=
1/N \frac{\partial }{\partial \beta} \left( \log \left(\left| \Delta C_{\gamma}\right|\right) - \sum_{\nu=1}^2 \log\left( \left| C^{(\nu)}_{\gamma}\right| \right) \right)
\\
&=\!\! 1/N  \left( \frac{1}{\left| \Delta C_{\gamma}\right|} \frac{\partial }{\partial \beta} \left| \Delta C_{\gamma}\right| - \sum_{\nu=1}^2  \frac{1}{\left| C^{(\nu)}_{\gamma}\right|}  \frac{\partial }{\partial \beta} \left|C^{(\nu)}_{\gamma}\right|  \right)
\end{align}
\begin{align}
&=\!\! 1/N  \!\!\left( \sum_{\nu=1}^2 \!\!\left(
\frac{\!\!\int_{-\infty}^{\infty}\!\!\!R_{\nu} \exp\left[ -\beta R_{\nu} \right] {d\mathbf{U}}}{\!\!\int_{-\infty}^{\infty}\!\!\!\exp\left[ -\beta R_{\nu} \right] {d\mathbf{U}}} 
\right)\!\!
- 
\frac{\!\!\int_{-\infty}^{\infty}\!\!\!R_{\Delta} \exp\left[ -\beta R_{\Delta} \right] {d\mathbf{U}}}{\!\!\int_{-\infty}^{\infty}\!\!\!\exp\left[ -\beta R_{\Delta} \right] {d\mathbf{U}}}  
  \right) \nonumber
 \\
&=\!\!
1/N  \left( \sum_{\nu=1}^2 \mathbb{E}[R_{\nu}]_{p_G(R_{\nu})} - \mathbb{E}[R_{\Delta}]_{p_G(R_{\Delta})} \right)
\end{align}
Where $p_G(R_{\Delta})=Z^{-1}\exp(-\beta R_{\Delta})$ is the Gibbs distribution with normalization constant $Z=\int_{-\infty}^{\infty}\!\!\!\exp\left[ -\beta R_{\Delta} \right] {d\mathbf{U}}$.
These expectations can easily be computed either for a finite set of transformations or with a continuous integral. 

Accordingly,  the second derivative is:
\begin{align}
&\frac{\partial^2 I(\beta)}{\partial \beta^2}
=
1/N  \left( \sum_{\nu=1}^2 \left( \frac{\partial }{\partial \beta}
\frac{\!\!\int_{-\infty}^{\infty}\!\!\!R_{\nu} \exp\left[ -\beta R_{\nu} \right] {d\mathbf{U}}}{\!\!\int_{-\infty}^{\infty}\!\!\!\exp\left[ -\beta R_{\nu} \right] {d\mathbf{U}}} 
\right)  \right. \nonumber
\\
&\quad\quad\quad
\left.- 
\frac{\partial }{\partial \beta}
\frac{\!\!\int_{-\infty}^{\infty}\!\!\!R_{\Delta} \exp\left[ -\beta R_{\Delta} \right] {d\mathbf{U}}}{\!\!\int_{-\infty}^{\infty}\!\!\!\exp\left[ -\beta R_{\Delta} \right] {d\mathbf{U}}}  
  \right)
  \\
&=\!\!
1/N  \left( \sum_{\nu=1}^2 \left( \mathbb{E}[R_{\nu}]_{p_G(R_{\nu})}^2 - \mathbb{E}[R^2_{\nu}]_{p_G(R_{\nu})} \right)  \right)
 \nonumber
\\
 &\quad\quad\quad -1/N   \left(\mathbb{E}[R^2_{\Delta}]_{p_G(R_{\Delta})} - \mathbb{E}[R_{\Delta}]_{p_G(R_{\Delta})}^2 \right)
  \end{align}

 \begin{widetext}

 \section*{Appendix B: Analytical Calculation with bounded integration range}
 
 We derive the mutual information 
 Eq.~(\ref{mInfo_analyt_bounded}) when the transformations are weighted with a Gaussian centered around the identity transformation $u^{(1)}_{it}$. Except for this modification the derivation is analog to the derivation of the unconstraint mutual information in Eq.~(\ref{mInfo_analyt}).
 The approximation sets and the joint approximation set are
 \begin{eqnarray}
\left| C^{(\nu)}_{\gamma}\right| &=& \int_{-\infty}^{\infty} \exp\left( -\beta R(\mathbf{X}^{(\nu)},\mathbf{U},\mathbf{S}^{(\nu)},\mathbf{V}^{(\nu)} ) \right)  \exp\left(-\frac{1}{2\sigma}\sum_{i,t} (u_{it}-u^{(1)}_{it})^2\right)  {d\mathbf{U}} \ ,\nu\in\left\{1,2\right\}
\\
&=&  \prod_{ij} \exp\left(-\beta {x_{ij}^{(\nu)}}^2 \right)  \exp\left( -\frac{1}{2\sigma D} \sum_{t}u_{it}^{*2} \right)  \\
&& \quad \cdot
\int_{-\infty}^{\infty} \exp\left( 
-\beta  \sum_t^k u_{it}\left(   \frac{1}{\sigma\beta D} u^{(1)}_{it} -2x_{ij}^{(\nu)}w_{tj}^{(\nu)} \right) \right)
\nonumber
\\
&& \quad \quad \quad \: \cdot
\exp\left( 
-2\beta   \sum_t^k u_{it} \left( w_{tj}^{(\nu)2}u_{it} + 2w_{tj}^{(\nu)}\sum_{t'\neq t}^k u_{it'}w_{t'j}^{(\nu)} +\frac{1}{2\sigma\beta D}u_{it} \right)       \right)  {d^k\mathbf{u_{i,.}}}  \nonumber
 \end{eqnarray}
 and
 \begin{eqnarray}
\left| \Delta C_{\gamma}\right| &=& \int_{-\infty}^{\infty} \exp\left( -\beta/2 \left( R(\mathbf{X}^{(1)},\mathbf{U},\mathbf{S}^{(1)},\mathbf{V}^{(1)} )
+
R(\mathbf{X}^{(2)},\mathbf{U},\mathbf{S}^{(1)},\mathbf{V}^{(1)} )
\right) \right)  \exp\left(-\frac{1}{2\sigma}\sum_{i,t} (u_{it}-u^{(1)}_{it})^2\right)  {d\mathbf{U}} 
\\
&=&\prod_{ij} \exp\left(-\beta/2 \left({x_{ij}^{(1)}}^2 +{x_{ij}^{(2)}}^2 \right) \right) \exp\left( -\frac{1}{2\sigma D} \sum_{t}u_{it}^{*2} \right)  
\\
&& 
\quad \cdot
\int_{-\infty}^{\infty} \exp\left( -\beta  \sum_t^k u_{it} \left( w_{tj}^{(1)2}u_{it} + 2w_{tj}^{(1)}\sum_{t'\neq t}^k u_{it'}w_{t'j}^{(1)} +\frac{1}{2\sigma\beta D}u_{it}   \right) \right)
\nonumber
 \\
&& \quad \quad \quad  \: \cdot
\exp\left(-\beta\sum_t^k u_{it}\left( (x_{ij}^{(1)}+x_{ij}^{(2)})w_{tj}^{(1)} - \frac{1}{\sigma\beta D} u^{(1)}_{it} \right)     \right)  {d^k\mathbf{u_{i,.}}} \ .
\nonumber
\end{eqnarray}
The characteristic terms of the integrals are
\begin{align}
&\mathbf{A^{(\nu)}_{(j)}} = 2\beta \mathbf{a^{(\nu)}_{(j)}} ,
\\
&\mathbf{b^{(\nu)}_{(i,j)}} =2\beta \;x_{ij}^{(\nu)}\mathbf{w_{.,j}^{(\nu)} }^T  -\frac{1}{\sigma D}\mathbf{u^{(1)}_{i,.}} \ ,
\\
&\mathbf{A^{(\Delta)}_{(j)}} = \mathbf{A^{(1)}_{(j)}} \ ,
\\
&\mathbf{b^{(\Delta)}_{(i,j)}} = \beta \;\left(x_{ij}^{(1)}+x_{ij}^{(2)}\right)\mathbf{w_{.,j}^{(1)} }^T -\frac{1}{\sigma D}\mathbf{u^{(1)}_{i,.}}\ ,
\end{align}
with
\begin{align}
&\mathbf{a^{(\nu)}_{(j)}}   :=  \mathbf{w_{.,j}^{(\nu)} }^T\mathbf{w_{.,j}^{(\nu)} }   + \frac{1}{2\sigma\beta D} \mathbb{I} \ .
\end{align}

\noindent
These terms determine the cardinalities of the approximation sets:
\begin{eqnarray}
\left| C^{(\nu)}_{\gamma}\right| &=& 
 \prod_{i,j} \exp\left(-\beta x_{ij}^{(\nu)2}\right) \exp\left(-\frac{1}{2\sigma D} \mathbf{u^{(1)}_{i,.}}\mathbf{u^{(1)}_{i,.}}^T \right)
\sqrt{ \frac{(2\pi)^k}{\det\left( 2\beta \mathbf{a^{(\nu)}_{(j)}}   \right) }   }  
 \exp\left(\frac{1}{2} \mathbf{b^{(\nu)}_{(i,j)}}^T\mathbf{A^{(\nu)}_{(j)}}^{-1}\mathbf{b^{(\nu)}_{(i,j)}}\right)
\\
&=&   
\left(\frac{\pi}{\beta}\right)^{\frac{NDk}{2}} \!\! \prod_{i,j} \det\left( \mathbf{a^{(\nu)}_{(j)}} \right)^{-1/2}  
\\
&&
 \!\!\exp\left(-\beta x_{ij}^{(\nu)2} -\frac{1}{2\sigma D} \mathbf{u^{(1)}_{i,.}}\mathbf{u^{(1)}_{i,.}}^T   + \frac{1}{4 \beta} 
		(2\beta \;x_{ij}^{(\nu)}\mathbf{w_{.,j}^{(\nu)} }^T  -\frac{1}{\sigma D}\mathbf{u^{(1)}_{i,.}})^T   \mathbf{a^{(\nu)}_{(j)}}^{-1} 
		(2\beta \;x_{ij}^{(\nu)}\mathbf{w_{.,j}^{(\nu)} }^T  -\frac{1}{\sigma D}\mathbf{u^{(1)}_{i,.}}) \right)
		\nonumber
\\
&=&   
\left(\frac{\pi}{\beta}\right)^{\frac{NDk}{2}} 
\left(  \prod_{j} \det\left( \mathbf{a^{(\nu)}_{(j)}} \right) \right)^{-N/2}
\\
&&
\cdot \prod_{i,j} \exp\left(-\beta x_{ij}^{(\nu)2}\left(1- \mathbf{w_{.,j}^{(\nu)} } \mathbf{a^{(\nu)}_{(j)}}^{-1} \mathbf{w_{.,j}^{(\nu)} }^T  \right)  
  		-\frac{1}{2\sigma D} \mathbf{u^{(1)}_{i,.}}\mathbf{u^{(1)}_{i,.}}^T		
		- \frac{1}{4 \beta \sigma^2 D^2} \mathbf{u^{(1)}_{i,.}} \mathbf{a^{(\nu)}_{(j)}}^{-1}  \mathbf{u^{(1)}_{i,.}}^T	
		\right) \nonumber
\end{eqnarray}
and
\begin{align}
\left| \Delta C_{\gamma}\right|
&\!\!= \!\! \prod_{i,j} \exp\left(-\frac{\beta}{2} \left(x_{ij}^{(1)2}+x_{ij}^{(2)2} \right) \right) \exp\left(-\frac{1}{2\sigma D} \mathbf{u^{(1)}_{i,.}}\mathbf{u^{(1)}_{i,.}}^T \right)
\sqrt{ \frac{(2\pi)^k}{\det\left( 2\beta \mathbf{a^{(1)}_{(j)}}   \right) }   }  
 \exp\left(\frac{1}{2} \mathbf{b^{(\Delta)}_{(i,j)}}^T\mathbf{A^{(1)}_{(j)}}^{-1}\mathbf{b^{(\Delta)}_{(i,j)}}\right)
\\
&\!\!=  
\!\!\left(\frac{\pi}{\beta}\right)^{\frac{NDk}{2}} \!\!\!\prod_{j} \det\left( \mathbf{a^{(1)}_{(j)}} \right)^{-N/2}  %
\\
&
 \prod_{i,j} \exp\left(-\frac{\beta}{2} 
\left(x_{ij}^{(1)2}+x_{ij}^{(2)2}   - \frac{1}{2}\left(x_{ij}^{(1)}+x_{ij}^{(2)} \right)^2 
\mathbf{w_{.,j}^{(1)} } \mathbf{a^{(1)}_{(j)}}^{-1} \mathbf{w_{.,j}^{(1)} }^T \right)
  		-\frac{1}{2\sigma D} \mathbf{u^{(1)}_{i,.}}\mathbf{u^{(1)}_{i,.}}^T		
		\!- \frac{1}{4 \beta \sigma^2 D^2} \mathbf{u^{(1)}_{i,.}} \mathbf{a^{(1)}_{(j)}}^{-1}  \mathbf{u^{(1)}_{i,.}}^T	
		\right) \nonumber
\end{align}

\noindent
The number of random transformations depends on the size of the Gaussian sphere $b$ :
\begin{eqnarray}
b &=& \int_{-\infty}^{\infty} \exp\left(-\frac{1}{2\sigma}\sum_{i,t} (u_{it}-u^{(1)}_{it})^2\right)  {d\mathbf{U}} 
\\
  &=&   \prod_{i} \sqrt{ \frac{(2\pi)^k}{\det\left( \sigma^{-1} \mathbb{I}  \right) }   } 
  \\
  &=&    
  \left( 2\pi\sigma \right)^{\frac{Nk}{2}}
  \ .
\end{eqnarray}

\noindent
Define the auxiliary variables $\mathbf{F^{(\nu)}_{(j)}}:=\mathbf{w_{.,j}^{(\nu)} } \mathbf{a^{(\nu)}_{(s)}}^{-1} \mathbf{w_{.,j}^{(\nu)} }^T$. Then, the mutual information is 
 \begin{eqnarray}
I(\beta) 
&=& 
1/N \left( \left|\left\{ \tau_j \right\}\right| + \log \left(\left| \Delta C_{\gamma}\right|\right) - \sum_{\nu=1}^2 \log\left( \left| C^{(\nu)}_{\gamma}\right| \right) \right)
\\
&=& 
\frac{k}{2}\log(2\pi\sigma) + \frac{Dk}{2} \left(\log\left(\beta\right)-\log\left(\pi\right)\right)
+ \frac{1}{2} \sum_j  \log\left(\det\left(\mathbf{a^{(2)}_{(j)}} \right)\right)
\\
&& 
-  \frac{1}{N}\sum_{ij}\left[ \frac{\beta}{2}  
\left(x_{ij}^{(1)2}+x_{ij}^{(2)2}   - \frac{1}{2}\left(x_{ij}^{(1)}+x_{ij}^{(2)} \right)^2 \mathbf{F^{(1)}_{(j)}} \right)   
	+ \frac{1}{2\sigma D}  \mathbf{u^{(1)}_{i,.}}\mathbf{u^{(1)}_{i,.}}^T		
		+ \frac{1}{4 \beta \sigma^2 D^2}  \mathbf{u^{(1)}_{i,.}} \mathbf{a^{(1)}_{(j)}}^{-1}  \mathbf{u^{(1)}_{i,.}}^T	\right]
	\nonumber
		\\
&&
+  \frac{1}{N}\sum_{ij} \left[ \frac{\beta}{2} x_{ij}^{(1)2}\left(2- 2\mathbf{F^{(1)}_{(j)}}  \right)  
  		+\frac{1}{2\sigma D} \mathbf{u^{(1)}_{i,.}}\mathbf{u^{(1)}_{i,.}}^T		
		+\frac{1}{4 \beta \sigma^2 D^2} \mathbf{u^{(1)}_{i,.}} \mathbf{a^{(1)}_{(j)}}^{-1}  \mathbf{u^{(1)}_{i,.}}^T	\right]
	\nonumber
			\\
&&
+  \frac{1}{N}\sum_{ij} \left[ \frac{\beta}{2} x_{ij}^{(2)2}\left(2- 2\mathbf{F^{(2)}_{(j)}}  \right)  
  		+\frac{1}{2\sigma D} \mathbf{u^{(1)}_{i,.}}\mathbf{u^{(1)}_{i,.}}^T		
		+\frac{1}{4 \beta \sigma^2 D^2} \mathbf{u^{(1)}_{i,.}} \mathbf{a^{(2)}_{(j)}}^{-1}  \mathbf{u^{(1)}_{i,.}}^T	\right]
\nonumber
	\\
&=& 
  \frac{k}{2}\log(2\pi\sigma) +\frac{kD}{2}\log\left(\frac{\beta}{\pi}\right)
+ \frac{1}{2} \sum_j  \log\left(\det\left(\mathbf{a^{(2)}_{(j)}} \right)\right)
\\
&&
+ \frac{1}{2\sigma N } \sum_{i} \mathbf{u^{(1)}_{i,.}}\mathbf{u^{(1)}_{i,.}}^T    
+ \frac{1}{4 \beta \sigma^2 D^2 N} \sum_{ij} \mathbf{u^{(1)}_{i,.}} \mathbf{a^{(2)}_{(j)}}^{-1}  \mathbf{u^{(1)}_{i,.}}^T
\nonumber
	\\
&&
+ \frac{\beta}{2N} \sum_{ij} \left[  x_{ij}^{(1)2}\left(1- \frac{3}{2}\mathbf{F^{(1)}_{(j)}}  \right)  
+  x_{ij}^{(2)2}\left(1 -2\mathbf{F^{(2)}_{(j)}}      +\frac{1}{2}\mathbf{F^{(1)}_{(j)}}  \right) + x_{ij}^{(1)}x_{ij}^{(2)} \mathbf{F^{(1)}_{(j)}}\right] 
 \label{mInfo_analyt_bounded}
	\ . 
\end{eqnarray}

\end{widetext}

$\ $

\end{document}